\documentclass{emulateapj}
\usepackage{amsmath}
\usepackage{color}

\newcommand{\myemail}{kzk15@psu.edu}

\slugcomment{}

\shorttitle{HE $\nu$ and $\gamma$ transients from trans-relativistic SN shock breakouts}
\shortauthors{Kashiyama et al.}

\begin{document}

\title{High energy neutrino and gamma-ray transients from \\ trans-relativistic supernova shock breakouts}

\author{Kazumi Kashiyama\altaffilmark{1}, Kohta Murase\altaffilmark{2,3}, 
Shunsaku Horiuchi\altaffilmark{2,4}, Shan Gao\altaffilmark{1}, and Peter
M\'esz\'aros\altaffilmark{1} }
\email{\myemail}

\altaffiltext{1}{Department of Astronomy \& Astrophysics; Department of Physics; Center for Particle \& Gravitational Astrophysics; Pennsylvania State University, University Park, PA 16802}
\altaffiltext{2}{CCAPP \& Department of Physics, Ohio State University, 191 W. Woodruff Ave., Columbus, Ohio 43210}
\altaffiltext{3}{Hubble Fellow, School of Natural Sciences, Institute for Advanced Study, 1 Einstein Dr. Princeton NJ 08540}
\altaffiltext{4}{Center for Cosmology, Department of Physics and Astronomy, University of California, Irvine, CA 92697-4575, USA}

\begin{abstract}
Trans-relativistic shocks that accompany some supernovae (SNe) produce X-ray burst emissions as they break out in the dense circumstellar medium around the progenitors.
This phenomenon is sometimes associated with peculiar low-luminosity gamma-ray bursts (LL GRBs). 
Here, we investigate the high energy neutrino and gamma-ray counterparts of such a class of SNe. 
Just beyond the shock breakout radius, particle acceleration in the collisionless shock starts to operate in the presence of breakout photons.  
We show that protons may be accelerated to sufficiently high energies and produce high energy neutrinos and gamma rays via the photomeson interaction. 
These neutrinos and gamma rays may be detectable from $\lesssim 10$~Mpc away by IceCube/KM3Net as multi-TeV transients almost simultaneously with the X-ray breakout, 
and even from $\lesssim 100$~Mpc away with follow-up observations by CTA using a wide-field sky monitor like {\it Swift} as a trigger. 
A statistical technique using a stacking approach could also be possible for the detection, with the aid of the SN optical/infrared counterparts.
Such multi-messenger observations offer the possibility to probe the transition of trans-relativistic shocks from radiation-mediated to collisionless ones, 
and would also constrain the mechanisms of particle acceleration and emission in LL GRBs. 
\end{abstract}

\keywords{acceleration of particles --- neutrinos --- shock waves --- supernovae: general}

\section{Introduction}
\label{sec:intro}
There is increasing evidence for a class of supernovae (SNe) which are accompanied by trans-relativistic shocks ($\beta_{\rm sh} \Gamma_{\rm sh} \sim 1$).
These are initially radiation mediated while going through optically thick gas, {\it i.e.,} the stellar or a circumstellar envelope \citep{1976ApJS...32..233W}.  
The typical photon energy could be as high as $\sim 200$ keV within a few Thomson mean free paths near the shock \citep{2010ApJ...716..781K, 2010ApJ...725...63B,2012ApJ...747...88N}. 
Once the optical depth becomes small enough that the photon diffusion velocity is similar to the shock velocity, photons begin to escape and are observed as the shock breakout emission.  
The energy spectrum is expected to be a combination of non-thermal and quasi-thermal \citep[e.g.,][]{Wang+07breakout,Waxman+07breakout,2010ApJ...725..904N}.
Such X-ray breakouts have now been confirmed in some cases \citep{2008Natur.453..469S}.

An interesting application of trans-relativistic shock breakouts is to low luminosity gamma-ray bursts~(LL GRBs) associated with SNe,  
such as GRB060218/SN2006aj \citep{2006Natur.442.1008C,2006Natur.442.1014S} and GRB100316D/SN2010bh \citep{2010arXiv1004.2262C,2011MNRAS.411.2792S}. 
These GRBs share peculiar characteristics compared to typical long GRBs: smaller luminosities ($L_{\rm iso, \gamma} \sim 10^{46} \ {\rm erg/s}$), 
softer peak energies ($\varepsilon_{\rm peak} \sim 1\mbox{-}10 \ {\rm keV}$), and longer durations ($t_{\gamma} \sim 1000 \ {\rm sec}$).  
The trans-relativistic shock breakout model can explain these characteristics \citep{2006Natur.442.1014S,2012ApJ...747...88N}.
The host galaxies are both nearby ($z = 0.0331$ for GRB060218 \citep{2006ApJ...643L..99M} and $z=0.059$ for GRB100316D \citep{2010GCN..10512...1V} ). 
The dimness and the proximity indicate that there might be a considerable number of sub-threshold events 
resulting in a relatively high local rate of some $R_{\rm LL}(z=0) \lesssim 10^{2\mbox{-}3} \ {\rm Gpc}^{-3}{\rm yr}^{-1}$ \citep{2007ApJ...657L..73G,2007ApJ...662.1111L}.  
Hence LL GRBs may be one of the promising targets for multi-messenger astronomy searches \citep{2012arXiv1203.5192A}, 
e.g. with specially designed facilities such as AMON\citep{2012arXiv1211.5602S}.

In general, shocks would become collisionless beyond the breakout radii, allowing charged particles to be accelerated \citep{2011PhRvD..84d3003M,2011arXiv1106.1898K}. 
The accelerated protons can produce mesons by the photomeson interaction or the inelastic $pp$ reaction, resulting in neutrinos and gamma rays as decay products of the mesons. 
This situation is analogous to the photospheric scenario of GRB prompt emission \citep{2008PhRvD..78j1302M,2009ApJ...691L..67W,Gao+12photnu}.  

In this {\it Letter}, we evaluate the high energy neutrino and gamma-ray emission from trans-relativistic shocks breaking out from the circumstellar envelope. 
We are primarily concerned with the application to LL GRBs associated SNe, 
but our prescriptions could be generally applied to other cases, e.g., non-GRB broad-line SNIc including hypernovae, 
by appropriately scaling the velocity and the breakout radius of the shock, depending on the density profile.

Our work differs from previous studies \citep{2006ApJ...651L...5M,2007APh....27..386G} 
where neutrino counterparts of LL GRBs were evaluated based on the {\it relativistic} jet model ($\Gamma_{\rm j} \gg 1$), 
which is also viable \citep{Mazzali+06,2007ApJ...659.1420T,2007MNRAS.375L..36G}. 
Although the main microphysical process, {\it i.e.} the photomeson interaction between accelerated protons and the observed X-ray photons, is the same, 
the predictions are not the same, due to differences in the emission radius and the Lorentz factor \citep{2011arXiv1106.1898K}.
Also, our prediction on the detectability of the gamma-ray counterpart is more positive than previous estimates. 
The present results also differ from those of \cite{2011PhRvD..84d3003M}, 
which considered the neutrino and gamma-ray emission from {\it non-relativistic} shocks ($\beta_{\rm sh} \lesssim 0.01$) colliding with the dense circumstellar material. 

\section{Model Set-up}
\label{sec:model}
We consider a trans-relativistic shock $(\beta_{\rm sh} \Gamma_{\rm sh} \sim 1)$ in a wind circumstellar material $(\rho \propto r^{-2})$. 
Initially, the shock is radiation-mediated, the temperature near the shock can be as high as $\sim 200 \ {\rm keV}$, 
and $e^{\pm}$ pairs are relevant as an opacity source \citep[e.g.,][]{2010ApJ...725...63B}. 
This hot, optically thick region begins to expand until the pair loading becomes negligible, that is, the temperature is decreased to $\lesssim 50 \ {\rm keV}$ \citep{2012ApJ...747...88N}. 
We define the breakout radius where a bulk of the photons begin to diffuse out as $r_{\rm sb} = y_{\rm \pm}r_{\rm sb}^{\rm NR}$. 
Here $r_{\rm sb}^{\rm NR}$ is the breakout radius for a non-relativistic shock obtained from 
$r_{\rm sb}^{\rm NR} \approx m_{\rm p}/\sigma_{\rm T} \rho(r_{\rm sb}^{\rm NR}) \beta_{\rm sh}$, where pairs are absent. 
The factor $y_{\pm} \geq 1$ represents the effect of pair opacity, expected to be $O(1)$ for trans-relativistic shocks. 
The duration of the breakout emission is 
\begin{equation}
t_\gamma \approx r_{\rm sb}/c \sim 3.0 \times 10^3 r_{{\rm sb,}13.95} \ {\rm sec}.
\end{equation} 
The energy trapped in the breakout shell is estimated as ${\mathcal E}_{\rm iso} \approx 2 \pi \rho(r_{\rm sb}) r_{\rm sb}{}^3 \beta_{\rm sh}{}^2 c^2$. 
Around the breakout, the density is 
\begin{eqnarray}\label{Rho}
\rho(r_{\rm sb}) &\approx& y_{\pm}{}^{-2} \rho(r_{\rm sb}^{\rm NR}) \notag \\
&\sim& 5.6 \times 10^{-14} \ y_{\pm}{}^{-1} r_{{\rm sb},13.95}{}^{-1} \beta_{\rm sh, 0.5}{}^{-1} \ {\rm g/cm^3}. 
\end{eqnarray}
Then, we have
\begin{equation}\label{E_iso}
{\mathcal E}_{\rm iso} \sim 5.8 \times 10^{49} \ y_{\pm}{}^{-1} r_{{\rm sb},13.95}{}^{2} \beta_{\rm sh, 0.5} \ {\rm erg}.  
\end{equation}
The mean photon luminosity is  
\begin{eqnarray}\label{L_iso_x}
L_{\rm iso, \gamma}  &\approx& \epsilon_{\gamma} {\mathcal E}_{\rm iso}/t_{\gamma} \notag \\ 
&\sim& 1.9 \times 10^{46} \ y_{\pm}{}^{-1} \epsilon_{\gamma} r_{{\rm sb},13.95} \beta_{\rm sh, 0.5} \ {\rm erg/sec}, 
\end{eqnarray}
where $\epsilon_{\gamma}$ is the fraction of the energy delivered to radiation. 
By setting $\beta_{\rm sh} = 0.5$ and $r_{\rm sb} = 9.0 \times 10^{13} \ {\rm cm}$ with $y_{\pm} \sim 1$ and $\epsilon_\gamma \sim 1$, 
we can roughly reproduce the characteristic of the prompt X-ray emissions of GRB060218 and GRB100316 \citep{2012ApJ...747...88N}. 

It is difficult to evaluate the time-dependent energy spectrum of the photon field in-situ at $r \gtrsim r_{\rm sb}$ \citep[e.g.,][]{2010ApJ...725..904N}.
In principle, not only non-thermal protons which we discuss later, 
but also non-thermal electrons accelerated at the collisionless shock can affect the photon spectrum by e.g., the inverse Compton process.
Here, we substitute instead the observed spectrum which, over most of the prompt emission episode, can be fitted approximately by a broken power law, 
$dn/d\varepsilon = n_{\rm b} (\varepsilon/\varepsilon_{\rm b}){}^{-\alpha}$ for $\varepsilon_{\rm min} < \varepsilon < \varepsilon_{\rm b}$, 
and $dn/d\varepsilon = n_{\rm b} (\varepsilon/\varepsilon_{\rm b}){}^{-\beta}$ for $\varepsilon_{\rm b} < \varepsilon < \varepsilon_{\rm max}$. 
We set $\alpha = 1.42$, $\beta =2.48$, $\varepsilon_{\rm b} = 16 \ {\rm keV}$ and $L_{\rm iso, \gamma}^b$ that is the luminosity at $\varepsilon_{\rm b}$, 
following the observation of GRB100316D \citep{2011ApJ...726...32F}.  
Note that $L_{\rm iso, \gamma}^b$ is related to $L_{\rm iso, \gamma}$ via 
$U_{\gamma} \equiv \int^{\varepsilon_{\rm max}}_{\varepsilon_{\rm min}} \varepsilon (dn/d\varepsilon) d\varepsilon = L_{\rm iso, \gamma}/ 4\pi r_{\rm sb}{}^2 c$. 
The low energy cut-off is assumed to be $\varepsilon_{\rm min} = 1.0 \ {\rm keV}$ where the observed SED declines sharply. 
In fact, {\it Swift} UVOT identified a prompt UV-optical emission with a photon index harder than $3/2$ for GRB060218 \citep{2007MNRAS.375L..36G}, 
and gave only an upper limit for GRB100316D \citep{2011MNRAS.411.2792S}. 
These soft photons can affect the opacity of high energy gamma rays, as we show below. 
The high energy cut-off can be set as $\varepsilon_{\rm max} = 1.0 \ {\rm MeV}$ since the pair absorption would be crucial above this energy.

If the matter is provided by a stellar wind where $\rho \approx \dot{M}/4\pi v_{\rm w} r_{\rm sb}{}^2$, the mass ejection rate of Eq.(\ref{Rho}) may be as large as 
$\dot{M} \sim 0.09 \ v_{\rm w,9} y_{\pm}{}^{-1} r_{{\rm sb},13.95} \beta_{\rm sh, 0.5}{}^{-1} \ M_{\odot}/{\rm yr}$ for a wind velocity $v_{\rm w} \sim 10^9 \ v_{\rm w,9} \ {\rm cm/s}$. 
Such massive winds in the pre-collapse phase have been inferred especially for type IIn SN progenitors 
\citep{2010ApJ...709..856S,2011ApJ...730...34S,2012ApJ...755..110C,2012ApJ...750L...2C,2012ApJ...751...92R}.

\section{Proton acceleration}
\label{sec:acc}

\begin{figure}
\includegraphics[width=8cm]{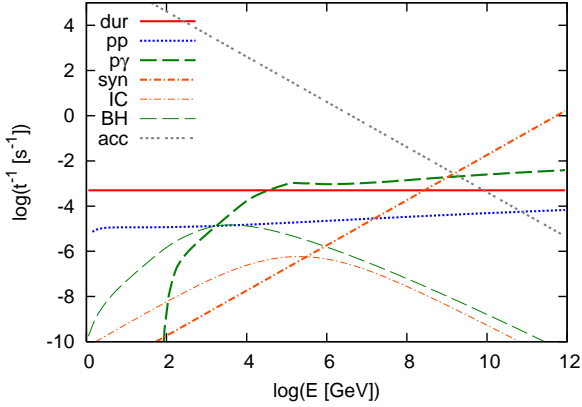}
\caption{
Acceleration timescale and cooling timescales of protons for a trans-relativistic shock breakout with $\beta_{\rm sh} = 0.5$, $r_{\rm sb} = 9 \times 10^{13} {\rm cm}$. 
For the breakout photons, we assume a broken power low with a bolometric luminosity $L_{\rm iso, \gamma}^b = 10^{46} \ {\rm erg/sec}$ as explained in the text. 
We also set $\xi_{\rm B} = 0.1$, and $\eta = 2 \pi$.
}
\label{f1}
\end{figure}

Beyond the breakout radius, the shock can no longer be mediated by the radiation. 
However, electromagnetic fields amplified by plasma instabilities will determine the shock structure, {\it i.e.} the shock becomes collisionless \citep{2011PhRvD..84d3003M,2011arXiv1106.1898K}. 
The magnetic field strength can be estimated as 
$B \approx (8 \pi \xi_{\rm B} U_{\gamma})^{1/2} \sim 4.0 \times 10^3 \ (\xi_{\rm B}/0.1)^{1/2} y_{\pm}{}^{-1/2} r_{{\rm sb},13.95}{}^{-1/2} \beta_{\rm sh, 0.5}{}^{1/2} \ {\rm G}$, 
where $\xi_{\rm B} \equiv U_{\rm B}/U_{\gamma}$ represents the magnetic field amplification efficiency. 
A fraction of the protons can be accelerated by the first-order Fermi process.
The maximum energy of the protons is determined by the condition, $t_{\rm acc} \leq t_{\rm p}$, 
where $t_{\rm acc} = \eta E_{\rm p}/eBc$ is the acceleration time (in the Bohm diffusion limit, $\eta \sim 2 \pi$), 
and $t_{\rm p}{}^{-1} \equiv t_{\rm \gamma}{}^{-1} + t_{\rm syn}{}^{-1} + t_{\rm IC}{}^{-1} + t_{\rm BH}{}^{-1} + t_{\rm p\gamma}{}^{-1} + t_{\rm pp}{}^{-1}$ is the total energy loss timescale, 
including the emission duration (dur) and the effect of the synchrotron radiation (syn), inverse Compton radiation (IC), the Bethe-Heitler type interactions (BH), 
the photomeson interactions (p$\rm \gamma$), and the inelastic pp collisions (pp). 
Fig.\ref{f1} shows the proton energy versus the various timescales, obtained with the calculation codes used in \cite{2008PhRvD..78j1302M}. 
One can see that the synchrotron cooling time $t_{\rm syn} = (3m_{\rm p}{}^4 c^3 / 4 \sigma_{\rm T} m_{\rm e}{}^2) U_{\rm B}{}^{-1} E_{\rm p}{}^{-1}$ 
is the most important for the proton maximum energy. 
The condition $t_{\rm acc} \lesssim t_{\rm syn}$ gives 
\begin{eqnarray}\label{eq_Ep_max}
E_{\rm p,max} &\lesssim& 7.3 \ \left( \frac{\xi_{\rm B}}{0.1} \right)^{-1/4} {\left( \frac{\eta}{2 \pi} \right)}^{-1/2} \notag \\ 
&\times& y_{\pm}{}^{1/4} r_{{\rm sb},13.95}{}^{1/4} \beta_{\rm sh, 0.5}{}^{-1/4} \ {\rm EeV}. 
\end{eqnarray}
We assume a power law distribution of the accelerated protons, $dN/dE_{\rm p} \propto E_{\rm p}^{-s}$ with $s = 2$. 
The peak fluxes of neutrinos and gamma-rays decrease by $\sim 30 \%$ for $s = 2.2$.
The normalization is determined by introducing the acceleration efficiency, $\epsilon_{\rm CR} \equiv {\mathcal E}_{\rm CR}/{\mathcal E}_{\rm iso}$ 
with ${\mathcal E}_{\rm CR} \equiv  \int^{E_{\rm p, max}} E_{\rm p} (dN/dE_{\rm p}) dE_{\rm p}$.  

\section{Neutrino and gamma-ray emission}
\label{sec:nu_gamma}

We consider the neutrinos and the gamma rays from the decay of mesons generated by both the photomeson production and inelastic pp reaction. 
In the analytical estimate below, we only discuss pions which turn out to give a dominant contribution.  
But the contribution from kaon decay is numerically included as in \cite{2008PhRvD..78j1302M}.

We can estimate the fraction of energy transferred from the non-thermal protons to the pions by the photomeson interactions as 
$\min [1, f_{\rm p\gamma}]$ where $f_{\rm p\gamma} \equiv  t_{\gamma}/t_{\rm p\gamma}$. 
Using the rectangular approximation \citep{1997PhRvL..78.2292W} for a photon spectrum approximated as a broken power law, we have 
\begin{eqnarray}\label{eq_fpg_D}
f_{\rm p\gamma} &\sim& y_{\pm}{}^{-1} \epsilon_{\gamma}  \varepsilon_{{\rm b,}16 {\rm keV}}{}^{-1} \notag  \\ 
&\times& 
\left\{\begin{array}{ll}
(E_{\rm p}/E_{\rm p,b})^{\beta-1} & \ \ \ \ (E_{\rm p} < E_{\rm p,b}), \\
(E_{\rm p}/E_{\rm p,b})^{\alpha-1} & \ \ \ \ (E_{\rm p,b} < E_{\rm p} ), \\
\end{array} \right.
\end{eqnarray}
where $E_{\rm p,b} =  0.5 \ \bar{\varepsilon}\varepsilon_{\rm b}{}^{-1} m_{\rm p}c^2 \sim 8.8 \ \varepsilon_{{\rm b, 16keV}}{}^{-1}$ TeV  with $\bar{\varepsilon} \sim 0.34 \ {\rm GeV}$. 
The multi-pion production becomes dominant above 
$\approx 0.5 \ \bar{\varepsilon}\varepsilon_{\rm min}{}^{-1} m_{\rm p}c^2 \sim 140 \ \varepsilon_{{\rm min, keV}}{}^{-1}$ TeV \citep[cf.][]{2008PhRvD..78b3005M}. 
We can conclude that a significant fraction of non-thermal protons with energies $10 \ {\rm TeV} \lesssim E_{\rm p} \lesssim {\rm EeV}$ will be converted into pions, 
even when $y_{\pm}$ is slightly larger than $1$.

The inelastic pp cooling time is $t_{\rm pp}{}^{-1} \approx (\rho/m_{\rm p})\kappa_{\rm pp} \sigma_{\rm pp} c$. 
The fraction of energy an incident proton loses, $f_{\rm pp} \equiv t_{\gamma}/t_{\rm pp}$, can be evaluated as 
\begin{equation}\label{eq_fpp}
f_{\rm pp} \sim 0.1 \ y_{\pm}{}^{-1} \beta_{\rm sh, 0.5}{}^{-2}, 
\end{equation}
where we use approximately constant values for the inelasticity $\kappa_{\rm pp} \sim 0.5$ 
and for the cross section $\sigma_{\rm pp} \sim 4 \times 10^{-26} \ {\rm cm^2}$, appropriate at high energies. 
Eq.(\ref{eq_fpp}) indicates that the inelastic pp collisions can also contribute moderately to the pion production as in the case of GRB photospheric emissions.

{\it Neutrino Emission.-}
Neutrinos are mainly produced as decay products of charged pions. 
One can find that the charged pions with $E_{\rm \pi} \gtrsim 5 \ (\xi_{\rm B}/0.1)^{-1/2} y_{\pm}{}^{1/2} \epsilon_{\gamma}{}^{-1/2} r_{{\rm sb},13.95}{}^{1/2} \beta_{\rm sh, 0.5}{}^{-1/2} \ {\rm PeV}$
will lose their energy before decaying due to the synchrotron cooling. 
Given that the resultant neutrinos have typically $\sim 1/4$ of the parent pion energy, one expects TeV-PeV neutrinos.
The peak fluence from a single supernova/burst event can be analytically estimated as 
$E_{\nu}{}^2 \phi_{\nu} \approx (1/4\pi D_{\rm L}^2) \times (1/4) \min[1,f_{\rm p\gamma}] (E_{\rm p}{}^2 dN/dE_{\rm p})$, or
\begin{eqnarray}\label{eq_ne_flu}
E_{\nu}{}^2 \phi_{\nu} &\sim& 10^{-5} \ \left(\frac{D_{\rm L}}{10 {\rm Mpc}}\right)^{-2}  \frac{\epsilon_{\rm CR}}{0.1} \notag \\ 
&\times& f_{\rm p\gamma} y_{\pm}{}^{-1} r_{{\rm sb},13.95}{}^2 \beta_{\rm sh, 0.5} \ {\rm erg/cm^{2}},
\end{eqnarray} 
where $D_{\rm L}$ is the luminosity distance to the source.

\begin{figure}
\includegraphics[width=8cm]{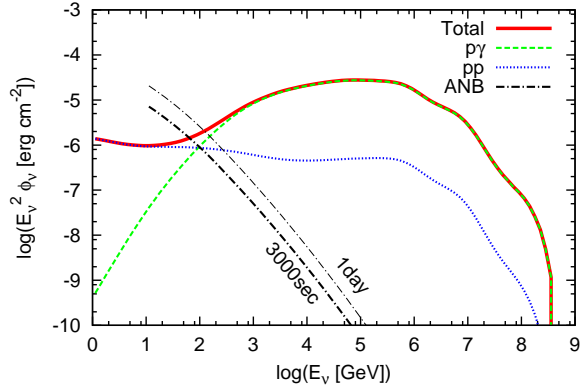}
\caption{
Energy fluences of neutrinos from a trans-relativistic shock breakout using the same parameters as Fig.\ref{f1}.  We set $\epsilon_{\rm CR} = 0.2$ and $D_{\rm L} = 10 \ {\rm Mpc}$. 
Lines represent a contribution from the photomeson production (dashed), the inelastic pp reaction (dotted), and the total (solid). 
The dotted-dashed lines show the zenith-angle averaged atmospheric neutrino background  (ANB)
within a circle of deg for $\Delta t = 3.0 \times 10^3 \ {\rm sec}$ (thick) and $\Delta t =  1$ day (thin). 
}
\label{f2}
\end{figure}

Fig.\ref{f2} shows the energy fluences of neutrinos obtained numerically using the calculation codes of \cite{2008PhRvD..78j1302M}, for the same parameters as Fig.\ref{f1}. 
The dashed and dotted lines show the contribution from the photomeson and inelastic pp interactions, respectively. 
We have verified that contributions from the kaon decay become important only above $\sim 10 \ {\rm PeV}$. 
The signal is above the zenith-angle-averaged atmospheric neutrino background (ANB, dotted-dash lines; thick one for $t_{\gamma} \sim 3 \times 10^3 \ {\rm sec}$, and thin one for $1$ day).
The number of muon events due to the muon neutrinos above TeV energies can be estimated as
$N_{\mu} \sim 0.3 \ (\epsilon_{\rm CR}/0.2) (D_{\rm L}/10 {\rm Mpc})^{-2} y_{\pm}{}^{-1} r_{{\rm sb},13.95}{}^2 \beta_{\rm sh, 0.5}$ 
using IceCube/KM3net \citep{2010arXiv1003.5715K,2006NIMPA.567..457K}.
Based on our fiducial parameters, IceCube/KM3net can marginally detect a nearby source at $\lesssim 10 \ {\rm Mpc}$, 
although such events occur rarely {\it i.e.}, $\lesssim 0.002 \ {\rm yr}^{-1}$ for a local LL GRB event rate $R_{\rm LL}(z=0) \sim 500 \ {\rm Gpc}^{-3} {\rm yr}^{-1}$ \citep{2007ApJ...657L..73G}.

From Fig.\ref{f2}, one can see that the typical neutrino energy in the trans-relativistic shock breakout model is ${\rm TeV\mbox{-}PeV}$. 
By comparison, the relativistic jet models of LL GRBs predict higher energy ${\rm PeV\mbox{-}EeV}$ neutrinos \citep{2006ApJ...651L...5M, 2007APh....27..386G}. 
This difference is mainly because the shock breakout model involves a lower Lorentz factor and a stronger cooling of mesons. 
In a relativistic jet, the peak photon energy in the comoving frame is $\varepsilon'_{\rm b} = \varepsilon_{\rm b}/\Gamma_{\rm j}$, where $\Gamma_{\rm j} $ is the Lorentz factor of the jet. 
The typical energy of protons interacting with photons via the photomeson production is 
${E_{\rm p}}' \sim 0.5 \ \bar{\varepsilon}{\varepsilon_{\rm b}}'{}^{-1} m_{\rm p}c^2$. 
The resultant neutrino energy will be ${E_{\rm \nu}} \sim 0.05 \times {E_{\rm p}}'  \Gamma_{\rm j}$ in the observer frame, 
which is $100 \ (\Gamma_{\rm j}/10){}^2$ times larger than our model. 
Thus, high-energy neutrino observations can provide clues to the emission model of LL GRBs.

In principle, the shock velocity could be independently constrained through the neutrino spectroscopy. 
From Eq.(\ref{eq_fpg_D}) and (\ref{eq_fpp}), both $f_{\rm p\gamma}$ and $f_{\rm pp}$ are present irrespective of $r_{\rm sb}$, and only depend on $\beta_{\rm sh}$. 
The relative importance of photomeson to inelastic pp collisions directly affects the neutrino energy spectrum.  
In the case of trans-relativistic shocks, the spectrum will have a bumpy structure like Fig.\ref{f2}. 
On the other hand, slower shocks will produce relatively flat spectra because of efficient inelastic pp interactions (see e.g., \cite{2011PhRvD..84d3003M}).

{\it Gamma-ray counterparts.-}
Gamma rays are mainly injected by neutral meson decays. 
Since the neutral mesons do not suffer synchrotron cooling, the maximum energy of gamma rays can be as high as $\sim 10 \%$ of the parent protons, 
that is $\sim 100 \ {\rm PeV}$ in our fiducial case. 
At high energies above $\sim$ MeV, the $e^{\pm}$ pair production can attenuate the gamma-ray flux. 
In the emission region, we can roughly take into account the attenuation by $\approx 1/(1+\tau_{\gamma\gamma})$, 
where $\tau_{\gamma\gamma}$ is the $e^{\pm}$ pair production optical depth  \citep[e.g.,][]{2006ApJ...650.1004B}. 
The observed photon spectrum of GRB100316D is employed to calculate the optical depth of the emission region numerically, 
with a Rayleigh-Jeans tail below $\varepsilon_{\rm min} = 1\ {\rm keV}$. 
This would be reasonable, since the result is not affected much as long as the photon index there is harder than $1$. 
We also take into account the attenuation by the extragalactic background light \citep[EBL;][]{2004A&A...413..807K}.

\begin{figure}
\includegraphics[width=8cm]{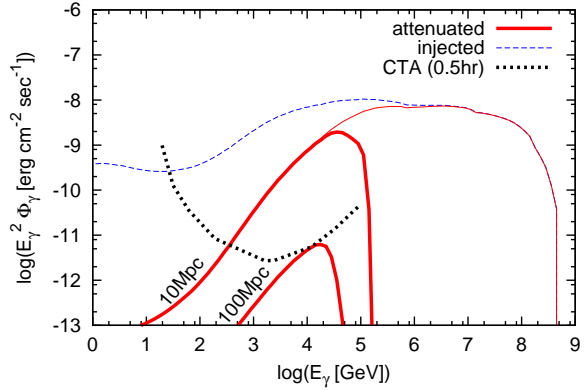}
\caption{
Energy fluxes of gamma rays corresponding to Fig.\ref{f2}.  
The emission duration is equal to that of X rays; $t_{\gamma} = 3.0 \times 10^3 \ {\rm sec}$. 
We show the cases of $D_{\rm L} = 10$ and $100 \ {\rm Mpc}$ (thick solid lines). 
For the former, we also show the injected spectrum without attenuation (dashed line) and only including attenuation within the emission region (thin line).
The dotted line shows $0.5 \ {\rm hr} = 1.8 \times 10^3 \ {\rm sec}$ differential sensitivity of CTA for a $5\sigma$ detection. 
}
\label{f3}
\end{figure}

Fig.\ref{f3} shows the numerically calculated energy spectrum of gamma rays.
The thick solid line represents the expected flux from a single LL GRB event at $10 \ {\rm Mpc}$ and $100 \ {\rm Mpc}$. 
The emission duration is set to that of the X rays, $t_{\gamma} \sim 3.0 \times 10^3 \ r_{{\rm sb},13.95} \ {\rm sec}$. 
As a reference, we also show the injected spectrum without attenuation (dashed line) and only including the attenuation within the emission region (thin solid line) for the $10 \ {\rm Mpc}$ case.
It can be seen that the attenuation of GeV $\lesssim E_{\rm \gamma} \lesssim 100$ TeV gamma rays 
is mainly due to the photon field in the emission region below/around $\varepsilon \gtrsim 1$ keV. 
In our case, the attenuation rate decreases with the energy because of the Klein-Nishina suppression. 
On the other hand, gamma rays above $\sim 100$ TeV are mostly attenuated by the EBL. 
In Fig.\ref{f3}, we also show the differential sensitivity of CTA for a $5\sigma$ detection with an exposure time comparable to $t_{\gamma}$, 
$0.5{\rm hr} = 1.8 \times 10^3 \ {\rm sec}$ \citep[dotted line;][]{2011ExA....32..193A}. 
CTA can detect the multi-TeV gamma rays even from $100 \ {\rm Mpc}$, 
within which the all-sky event rate would be $\sim 2 \ {\rm yr}^{-1}$ for $R_{\rm LL}(z=0) \sim 500 \ {\rm Gpc}^{-3} {\rm yr}^{-1}$.  
The FOV of CTA with the shown sensitivity $\sim  5 \ {\rm deg}$ would not be wide enough for a blind search.  
On the other hand, a survey mode with a wider FOV would not be sensitive enough to detect the signal. 
Thus, for CTA, a rapid follow-up observation triggered by a wide-field X-ray telescope such as {\it Swift} or a {\it Lobster}-type instrument is needed.  
Assuming that the sky coverage is $\gtrsim 10 \%$, one can expect $\gtrsim 0.2$ events ${\rm yr}^{-1}$ within $100$ Mpc.  
The detection rate would be increased by a simultaneous operation of HAWC 
with a sensitivity $\sim 10^{-10} {\rm erg \ cm^{-2} sec^{-1}}$ for $\sim 100$ TeV gamma rays \citep{2012NIMPA.692...72D}.

A detection of the multi-TeV gamma-ray transient, as expected in this model, would also constrain the emission mechanism of LL GRB. 
This is in contrast to the relativistic jet model, where as in the neutrino counterpart, the typical energy of the gamma rays injected by the photomeson reaction would be $\gtrsim$ PeV, 
which will be completely attenuated by the EBL even if they can escape the emission region.

\section{Summary and discussion} 
\label{sec:sum}
We have shown that trans-relativistic shock breakouts in SNe can be accompanied by multi-TeV neutrino and gamma-ray transients. 
These can provide diagnostics for a radiation mediated shock converting into a collisionless shock, and for baryon acceleration there. 
We can also get clues to the emission mechanism of LL GRBs by detecting such high energy counterparts simultaneously with the prompt X-ray emission. 
The multi-TeV gamma rays can be detectable even from $100$ Mpc away using CTA. 
These results motivate follow-up observations triggered by a wide-field X-ray telescope like {\it Swift}.

While typically one expects very few neutrino events from those trans-relativistic SNe,
nevertheless searches for them would be aided by other possible counterparts. 
Using the information of optical/infrared counterparts of core-collapse SNe, one can essentially fix the position within the angular resolution of IceCube/KM3net $\lesssim {\rm deg}$, 
and also restrict the time domain of the neutrino search within $\sim$ day. 
The atmospheric neutrino background (ANB) of IceCube/KM3net within a circle of a degree 
over a day is roughly $\lesssim 10^{-5} \ E_{\rm \nu, 100TeV}{}^{-2}$ events day$^{-1}$. 
In terms of this ANB flux, neutrinos from trans-relativistic shock breakouts within $D_{\rm L} \sim E_{\rm \nu, 100TeV} \ {\rm Gpc}$ can give a signal-to-noise ratio $\gtrsim 1$ (see also Fig.\ref{f2}). 
One could then statistically extract $O(1)$ astrophysical neutrinos by stacking the optical counterparts of $O(10^5)$ SNe within $z \lesssim 0.3$, 
whether or not the X-ray counterparts are observed.  
Given that the whole sky event rate of such LL GRBs would be $\sim 3 \times10^4$ yr$^{-1}$ assuming $R_{\rm LL}(z=0) \sim 500 \ {\rm Gpc}^{-3} {\rm yr}^{-1}$, 
a decadal SNe search up to $z \lesssim 0.3$ with a sky coverage $\gtrsim 10\%$ is needed. 
While still a challenging task, this kind of astronomy may be possible in the LSST era \citep{2009JCAP...01..047L}. 

Non-GRB broad-line SNIc can also be accompanied by trans-relativistic shocks with $\beta_{\rm sh} \Gamma_{\rm sh} \sim 1$, 
which break out at a certain radius and could produce neutrinos in the presence of material with a shallow profile.
Although the typical fraction of the energy in the trans-relativistic component would be relatively small, ${\mathcal E}_{\rm iso} \lesssim 10^{48} \ \rm erg$ \citep[e.g.,][]{2010Natur.463..513S},  
the event rate is larger than for LL GRBs, $R_{\rm Ibc}(z=0) \sim 2\times 10^3 \ {\rm Gpc}^{-3} {\rm yr}^{-1}$ \citep{1998MNRAS.297L..17M}.  
Thus, they could give a comparable or a larger amounts of neutrinos by using the above stacking method.

\acknowledgments
We thank Bin-Bin Zhang and Peter Veres for valuable discussions. This work is supported 
in part by a JSPS fellowship for research abroad, by NSF PHY-0757155 and the CCAPP 
workshop on "Revealing Deaths of Massive Stars with GeV-TeV Neutrinos".


\end{document}